\documentclass{bmcart}

\usepackage{natbib}
\usepackage{amsthm, amsmath, mathtools, amssymb}
\RequirePackage{natbib}
\RequirePackage{hyperref}
\usepackage[utf8]{inputenc} 
\usepackage{epstopdf}
\usepackage{graphicx}
\usepackage{epsfig}
\usepackage{algpseudocode}
\usepackage{algorithm}
\makeatletter
\def\BState{\State\hskip-\ALG@thistlm}
\makeatother

\startlocaldefs
\endlocaldefs

\begin{document}

\begin{frontmatter}

\begin{fmbox}
\dochead{Research}


\title{A Complex Networks Approach to Find Latent Clusters of Terrorist Groups}


\author[
   addressref={aff1},                   
   email={gianmaria.campedelli@unicatt.it}   
]{\inits{GM}\fnm{Gian Maria} \snm{Campedelli}}
\author[
   addressref={aff2},
   email={gianmaria.campedelli@unicatt.it}
]{\inits{IC}\fnm{Iain} \snm{Cruickshank}}
\author[
   addressref={aff2},
   email={icruicks@andrew.cmu.edu}
]{\inits{KM}\fnm{Kathleen M.} \snm{Carley}}


\address[id=aff1]{
  \orgname{Transcrime - Universit\`a Cattolica del Sacro Cuore}, 
  \street{L.go Gemelli, 1},                     %
  \city{Milan},                              
  \cny{Italy}                                    
}
\address[id=aff2]{%
  \orgname{School of Computer Science - Carnegie Mellon University},
  \street{5000 Forbes Avenue},
  \postcode{15213}
  \city{Pittsburgh, PA},
  \cny{USA}
}



\end{fmbox}


\begin{abstractbox}

\begin{abstract} 
 Given \color{black} the extreme heterogeneity of actors and groups participating in terrorist actions, investigating and assessing their characteristics can be important to extract relevant information and enhance the knowledge on their behaviors. The present work will seek to achieve this goal via a complex networks approach. This approach will allow finding latent clusters of similar terror groups using information on their operational characteristics. Specifically, using open access data of terrorist attacks occurred worldwide from 1997 to 2016, we build a multi-partite network that includes terrorist groups and related information on tactics, weapons, targets, active regions. We propose a novel algorithm for cluster formation that expands our earlier work that solely used Gower's coefficient of similarity via the application of Von Neumann entropy for mode-weighting.  This novel approach is compared with our previous Gower-based method and a heuristic clustering technique that only focuses on groups' ideologies. The comparative analysis demonstrates that the entropy-based approach tends to reliably reflect the structure of the data that naturally emerges from the baseline Gower-based method. Additionally, it provides interesting results in terms of behavioral and ideological characteristics of terrorist groups. We furthermore show that the ideology-based procedure tends to distort or hide existing patterns. Among the main statistical \color{black} results, our work reveals that groups belonging to opposite ideologies can share very common behaviors and that Islamist/jihadist groups hold peculiar behavioral characteristics with respect to the others. \color{black} Limitations and potential work directions are also discussed, introducing the idea of a dynamic entropy-based framework.
\end{abstract}




\end{abstractbox}
%

\end{frontmatter}



\section*{Introduction}
Complex networks have demonstrated their potential in many different domains. Approaches that rely on dynamic, multi-mode, multi-partite and meta-networks have been fruitful in shedding light on a wide variety of phenomena, including social ones \cite{BarabasiEvolutionsocialnetwork2002, CarleyComputationalorganizationscience2002, SzellMultirelationalorganizationlargescale2010, CentolaSpreadBehaviorOnline2010}. In the last years, this process has indeed also touched areas as criminology, international security, and terrorism research \cite{CranmerKantianfractionalizationpredicts2015, BerlusconiLinkPredictionCriminal2016, LiNetworksmodelEast2015} . 

This methodological shift has been facilitated by the increasing availability of open access data sets, the sensibility and interest of social scientists towards novel empirical approaches and the dramatic popularity of statistical software and data-science oriented languages. In spite of this shift, several critical points and pitfalls have been highlighted by scholars regarding the actual results of scientific inquiry in the field of terrorism research. Sageman \cite{sageman_stagnation_2014}, for instance, noted that the lacking collaboration between intelligence and academia led to a stagnation that is mainly motivated by the scarcity of rich, detailed and precise data on terrorist groups and events, which makes it difficult for researchers to develop models that are actually useful in reducing or assessing the terrorist threat.

In fact, Sageman argued that the intelligence community should be more willing to share crucial and rich data sets to the academia, in order to exploit their methodological rigour and capabilities. Recently, in an attempt to extensively review the field of terrorism research, Schuurman \cite{SchuurmanResearchTerrorism20072018a} noted that many longstanding weaknesses and issues have been either completely or partially solved (e.g. scholars have expanded the range of data gathering techniques), while at the same time other issues are still in place. Among the others, scarcity of international and interdisciplinary collaborations and the high number of one-time contributors are preventing the field to develop in a more structured direction, therefore limiting the probability for high-impact and practical research. 

In spite of these structural limitations, we seek to demonstrate the potential capabilities of complex networks to highlight hidden patterns within the terrorist global scenario, with the final aim to stimulate the debate on the application of novel methodological frameworks to research on terrorism. Hidden patterns could highlight operational similarities between groups that do not share any ideological background, peculiar attack-planning characteristics related to actors operating in certain areas, or even relevant behavioral differences between groups that fight for similar motivations but are settled in distinct regions. \color{black} Our intuition is that there is first and foremost a need for advancing and experimenting novel methodological approaches that, in case of promising results, might be employed and applied to other contexts with more reliable data, allowing to draw more useful and solid conclusions. In fact, the unavailability of and search for better data should not stop the process of innovation within the field. This works relies on data retrieved from the Global Terrorism Database (GTD henceforth) on terrorist attacks occurred at global level from 1997 to 2016. The paper aims to propose a new algorithm for detecting latent clusters of  terrorist groups expanding and extending the analytic approach we have provided in \cite{CampedelliDetectingLatentTerrorist2019}: we demonstrate that, tested against our previous approach (which we will refer to as ``baseline'' throughout the paper) and a weak heuristic approach based on pure groups ideology, our novel algorithm confirms many results obtained with the baseline approach and also provides new interesting results on the hidden similarities across groups belonging to very different contexts and motivations.

The paper is organized as follows: the next section provides a review of network- and clustering-based approaches to the study of terrorism, trying to identify the main areas of application in which these methods have been experimented. Following, the ``Data'' section will thoroughly present the information contained in the data employed to conduct our analyses. The ``Methodology'' section will describe and explain the graph construction and algorithmic framework. The ``Results'' section will then describe relevant outcomes and patterns found after the experiments. Finally, ``Discussion and Future Work'' section will focus on the potential implications of this work, on its limits and on the possible directions that can be explored using this paper as a starting point. 

\section*{Background}

In recent years, one of the methodological frameworks that have been tested and have attracted the attention of both scholars and policymakers in the social sciences is network science, broadly intended \citep{BorgattiNetworkAnalysisSocial2009a}.  Network science has gained popularity in sociology \citep{GranovetterSociologyEconomicLife2018, KeuschniggAnalyticalsociologycomputational2018, CentolaHowBehaviorSpreads2018}, economics \citep{SchweitzerEconomicNetworksNew2009, Hausmannnetworkstructureeconomic2011, VitaliNetworkGlobalCorporate2011, CaccioliNetworkmodelsfinancial2018}, political science \citep{Haysspatialmodelincorporating2010, WardNetworkAnalysisPolitical2011, GerberPoliticalHomophilyCollaboration2013, Ribeirodynamicalstructurepolitical2018} and criminology \citep{MorselliCriminalNetworks2009, PapachristosCompanyYouKeep2015, AgresteNetworkstructureresilience2016, CalderoniCommunitiescriminalnetworks2017, daCunhaTopologyrobustnessstructural2018}.  

\color{black} This also applies to terrorism research. The first application of social network analysis to terrorism, the branch of network science that specifically seeks to map and study relation between human entities such as people or organizations, was the renowned paper by Krebs \cite{krebs_mapping_2002}. Krebs tried to understand the existing connections between hijackers and terrorists responsible for the 9/11 attacks using unstructured data retrieved from open access sources, as newspapers. Despite its limited sophistication, the study opened a path towards the study of terrorism under this new perspective. Following this strategy, other scholars have used relational data on individuals to reconstruct terrorist networks and investigate roles and key players \cite{koschade_social_2006, BramsInfluenceTerroristNetworks2006, belli_exploring_2015}.

Shifting from the pure physical and relational information gathered and structured to investigate the structure of groups, scholars have also tested and simulated the strength and resilience of terrorist networks. As an example, works in this area have put a strong emphasis on the application for intelligence purposes. They have relied on mathematical models focusing on network topology for either proposing methods for maximizing efficiency in disruption strategies \cite{carley_destabilization_2006, LindelaufUnderstandingTerroristNetwork2011, EiseltDestabilizationterroristnetworks2018, RenGeneralizednetworkdismantling2019a} or understand the most resilient topology structures to be learned from terrorism behavior and applied to other domains (e.g. infrastructure networks) \cite{GutfraindOptimizingTopologicalCascade2010a}.
This interest towards increasingly complex questions regarding the nature and behavior of terrorist networks encourages scholars and scientists to integrate relational and topological information on networks with their spatial and temporal dynamics. Spatial and temporal dynamics are crucial when aiming at understanding the evolution of a certain entity or phenomenon, therefore several works have focused on these aspects using either synthetic-generated data or real-world information on existing networks \cite{moon_modeling_2007, MedinaAdvancingUnderstandingSociospatial2011}.

In the meanwhile, the revolution of social media has provided an unprecedented and massive amount of data to study the online social behaviour of people. As for the real physical world, individuals act criminally or violently also within the internet, and therefore researchers have started to be attracted by the potential consequences of criminal, and even terrorist, behaviors in the cyberspace. Indeed, a recent stream of research has focused on the detection of terrorist or radical behaviors retrieving network-information from social media platforms. Social media allow to  go beyond pure relational information, integrating instead geographical, temporal features and many other profile attributes to infer patterns and dynamics of extremist users \cite{BouchardPreliminaryAnalyticalConsiderations2014, ChatfieldTweetingpropagandaradicalization2015, klausen_tweeting_2015, benigni_online_2017}.

The previous lines of research, though generally different in their data gathering techniques, modelling architectures and complexity scales, all mostly focus on mapping relations between individuals or, at most, organizations belonging to the same terrorist sphere (e.g. the al Qaeda network). However, a very recent sub-domain explored the power of the complex network landscape when dealing with event data and abstract meta-networks of attack characteristics, with the aim to predict future terrorist behaviours in terms of target or weapon selection, targeted locations and employed tactics \cite{desmarais_forecasting_2013, TutunNewframeworkthat2017a, thomson_complex_2018} or, more broadly, to highlight operational similarities between different terrorist organizations \cite{CampedelliDetectingLatentTerrorist2019, CampedelliPairwisesimilarityjihadist2019}.  

While network approaches for modelling terrorism have gained a certain degree of success and have tested and experimented techniques focusing on a variety of research questions, it is worth to note how this advancements have not been followed by the consequent combination of network science with unsupervised learning and, more specifically, cluster analysis. In one of the first attempts at using cluster analysis to group terrorist organizations, Chenoweth and Lowham \cite{chenoweth_classifying_2007} used data on groups which targeted American citizens to explore alternative ways to conceive terrorist typologies. Qi et al. \cite{qi_hierarchical_2010} used both social network analysis and unsupervised learning to group extremist web pages using an hierarchical multi-membership clustering algorithm based on the similarity score of these pages. Finally, Lautenschlager et al \cite{lautenschlager_group_2015} developed the Group Profiling Automation for Crime and Terrorism (GPACT) prototype that generates terrorist group profiling via a multi-step methodology that also includes clustering of terrorist events. 

In light of this gap in research, following the intuition that network science may provide rich insights on the terror phenomenon, we modify our previous proposed methodology to test the performance of an automatic weighting scheme for the Gower's coefficient of similarity based on von Neumann's Entropy to preserve intrinsic qualities of the data that already emerge from our baseline approach. 

\section*{Data}
This work relies on data retrieved from the Global Terrorism Database (GTD) \cite{lafree_introducing_2007, national_consortium_for_the_study_of_terrorism_and_responses_to_terrorism_global_2016}. The GTD is the most comprehensive and detailed open access dataset on terrorist events at global scale, maintained by the START research center.
Information are gathered from different open sources, and events have to meet specific criteria to be included in the database. These criteria are divided into two different levels. \\
The first level criteria are three and have all to be verified. These mandatory ones are related to (1) intentionality of the incident, (2) presence of violence (or immediate threat of violence) of the incident and (3) to the sub-national nature of terrorist actors.\\
The second level criteria are three and at least two of them must be respected. Second level criteria relate to (1) the specific political, economic, religious or social goal of each act, (2) the evidence of an intention to coerce, intimidate or convey messages to larger audiences than the immediate victims, (3) the context of action which has to be outside of legitimate warfare activities. Finally, although an event respects these two levels, an additional filtering mechanism (variable \textit{doubter}) controls for conflicting information or acts that may not be of exclusive terrorist nature \citep{STARTGTDCodebookInclusion2017}. For our analysis, we aggregated data (i.e., we did not separate by year or other time windows) from 1997 to 2016 on worldwide events and related perpetrators, excluding all the attacks which were of doubtful terrorist nature.\footnote{We have also excluded attacks from 1970 to 1996 because, as reported in the official GTD codebook, many variables on attacks occurred prior to 1997 were not available or sufficiently reliable.} \color{black} This methodological choice led from 106,114 events to a total of 88,513. Furthermore, we have removed all the events plotted by ``Unknown" actors. Considering the large amount of attacks with no identified perpetrator, we would have faced the risk of biased results. We have thus kept only attacks of clear terrorist nature with an identified author, accounting for a total of 41,456 events. 

The multi-partite network which has been created and employed for our study relied on six main terrorist dimensions, namely: Events ($N$=41,456), Groups ($N$=1,493), Targets ($N$=22)\footnote{Targets list includes: Abortion Related, Government (General), Private Citizens \& Property,	Business,	Religious Figures/Institutions,	Police,	Airports \& Aircraft,	Utilities,	Educational Institution,	Unknown,	Journalists \& Media,	Government (Diplomatic),	Other,	Military,	Telecommunication,	Tourists,	Terrorists/Non-State Militia,	Transportation,	NGO,	Violent Political Party,	Maritime,	Food or Water Supply.}, Weapons ($N$=13)\footnote{Weapons list includes: Incendiary,	Explosives/Bombs/Dynamite,	Firearms,	Unknown,	Melee,	Fake Weapons,	Chemical,	Other,	Sabotage Equipment,	Vehicle (not to include vehicle-borne explosives, i.e., car or truck bombs), Biological}, Tactics ($N$=9)\footnote{Tactics list includes: Facility/Infrastructure Attack,	Bombing/Explosion,	Armed Assault,	Unknown,	Assassination,	Hostage Taking (Kidnapping),	Unarmed Assault,	Hijacking,	Hostage Taking (Barricade Incident)} and operating Regions ($N$=12) \footnote{Operating Region list covers the entire world and specifically includes: North America,	South Asia,	Middle East \& North Africa,	Sub-Saharan Africa,	Western Europe,	Eastern Europe,	South America,	Southeast Asia,	East Asia,	Central America \& Caribbean,	Australasia \& Oceania,	Central Asia}. These dimensions have  been chosen because they represent the visible core of terrorist activity: the terror attack itself can indeed be represented by its perpetrator, the chosen target, the employed weapons and tactics and the geographic and political context in which it occurred. These variables are thus helpful in gathering a rich knowledge structure that will then be crucial for our methodology.

\color{black} In addition to this information which represent the basis of this work, other variables extracted from the GTD and other sources have been employed to detect and assess behavioral patterns of terrorist groups belonging to the same clusters \textit{ex post}. This information will include group-based attributes regarding terrorist activity such as ideology, success rate, suicide rate, fatality rate, casualty rate, multiplot rate, international rate and number of targeted countries. The ideology of each group has been mapped using existing information present in two open access data sets (Big Allied and Dangerous 1 and an extraction of Big Allied and Dangerous 2) when that information was available within those sources \citep{AsalBigAlliedDangerous2015}, and by exception from other qualitative open access information sources. \\
This mapping led to include seven ideology categories: (i) Islamist/Jihadist groups, (ii) Far Left/Anarchist/Communist (FL), (iii) Far Right/Racist/Nazi (FR), \color{black} (iv) Ethno-Nationalist, (v) Other/Unknown, (vi) Religious (Islam excluded), (vii) Animal-rights/Environmentalist. A given group may belong to more than one category at a time (e.g.: the Popular Front for the Liberation of Palestine which contains at the same time elements of Marxism and Nationalism) (Table \ref{desc_ide}). It is worth to specify that these are labels that aim at giving context regarding the main motivations and ideological positions of the groups. This of course does not imply that enviromentalism, for instance, has to be associated with terrorism \textit{per se}. These categories only mean that a given group that has plotted at least one attack included in the GTD had motivations and roots that can be matched with a given ideology. The same applies to left-wing or right-wing organizations: having a particular political position does not automatically qualify an existing entity (either a person or a group) as terrorist. However, there are diverted and extremist positions on both political sides that are tightly connected with groups and actors that have been responsible of terrorist acts.

\color{black}

\begin{table}[!h]
\centering
\caption{Descriptive Statistics of Group Ideologies}
\begin{tabular}{lcc}
\hline
\textbf{Ideology}        & \multicolumn{1}{l}{\textbf{N groups}} & \multicolumn{1}{l}{\textbf{Share}} \\ \hline\hline
Islamist/Jihadism           & 462                                   & \textit{0.31}                      \\ \hline
FL          & 271                                   & \textit{0.18}                      \\ \hline
Ethno/Nationalist        & 601                                   & \textit{0.40}                      \\ \hline
FR             & 50                                    & \textit{0.03}                      \\ \hline
Other/Unknown            & 206                                   & \textit{0.14}                      \\ \hline
Religion (No Islam)      & 45                                    & \textit{0.03}                      \\ \hline
Animal/Environmentalist  & 14                                    & \textit{0.01}                      \\ \hline\hline
\textbf{N of ideologies} & \textbf{N groups}                            & \textbf{Share}                     \\ \hline
1                        & 1347                                  & \textit{0.90}                      \\ \hline
2                        & 136                                   & \textit{0.09}                      \\ \hline
3                        & 10                                    & \textit{0.01}                      \\ \hline
\textbf{Total}           & 1493                                  & \textit{1.00}                                     \\ \hline
\end{tabular}
\label{desc_ide}
\end{table}

The  \textit{success share} is given by the ratio between the successful attacks and the total number of events attributed to a given group. The \textit{suicide share} maps the ratio of suicide attacks over the total number of events plotted by the same group. \textit{Fatality} and \textit{casualty} ratios are produced by the number of attacks with at least one dead victim (fatality) or one wounded victim (casualty) divided by the total number of events. The  \textit{international rate} is simply the ratio between attacks with some international features (e.g. logistic organization) and the total number of attacks. Finally, \textit{multiplot share} quantifies the share of attacks that were part of a coordinated strategy (e.g. 9/11 case), out of total attacks. All these variables seek to enrich the knowledge associated to each group and to understand whether the identified clusters highlight certain patterned and eventually unexpected behaviors (Table \ref{desc_behl}). 

\begin{table}[!h]
\centering
\caption{Group-based Attributes on Terrorist Activity - Descriptive Statistics}
\label{desc_behl}
\begin{tabular}{l c c c c c}
\hline
\textbf{}               & \textbf{Mean} & \textbf{St.Dev.} & \textbf{Median} & \textbf{Min} & \textbf{Max} \\ \hline\hline
Events                  & 27.76         & 205.81          & 2.00            & 1.00         & 5,634       \\ \hline
Success Share           & 0.90          & 0.24             & 1.00            & 0.00         & 1.00         \\ \hline
Suicide Share            & 0.03          & 0.14             & 0.00            & 0.00         & 1.00         \\ \hline
Fatality Ratio           & 2.75          & 8.55             & 0.50            & 0.00         & 170          \\ \hline
Casualty Ratio           & 7.88          & 23.34            & 1.67            & 0.00         & 385.29       \\ \hline
Multiplot Share          & 0.14          & 0.28             & 0.00            & 0.00         & 1.00         \\ \hline
International Share      & 0.29          & 0.41             & 0.00            & 0.00         & 1.00         \\ \hline
N Targeted Countries & 1.35          & 1.83             & 1.00            & 1.00         & 42          \\ \hline
\end{tabular}
\end{table}
\section*{Methodology}

At the general level, the entropy-based approach that is presented and analyzed in this work is structured as follows: (i) calculation of the weights of each mode using the graph entropy of that mode; (ii) computation of the weighted Gower's Coefficient of Similarity between each of the terrorist groups using the entropy as the weight; (iii) extraction of the latent network from the pairwise Gower's Coefficient similarities and analyze its structural and intra-cluster properties. The detailed process is described in the following subsections. 
\\
The entropy-based method will be then compared with the baseline model presented in \cite{CampedelliDetectingLatentTerrorist2019} and with a heuristic method. The baseline model uses a simplified version of Gower's method. In this simplified version, no weights are applied to the different modes and we only consider the natural structure of the data deriving from the affinity matrix that originates by the pairwise Gower's coefficient of similarity. In other words, instead of using the graph entropy of each mode as its weight, every mode is simply just given a weight of one. The heuristic method we use for comparison only uses groups' ideologies as the clustering criterion. In this method, we just use the dominant ideology of a particular terrorist group as its cluster label. So, if two groups share a dominant ideology, like Ethno/Nationalist, then they are in the same cluster. 

\color{black}
\subsection*{Entropy-based Gower's Method for Multi-partite Data}
\begin{figure}[h!]
\includegraphics[scale=0.32]{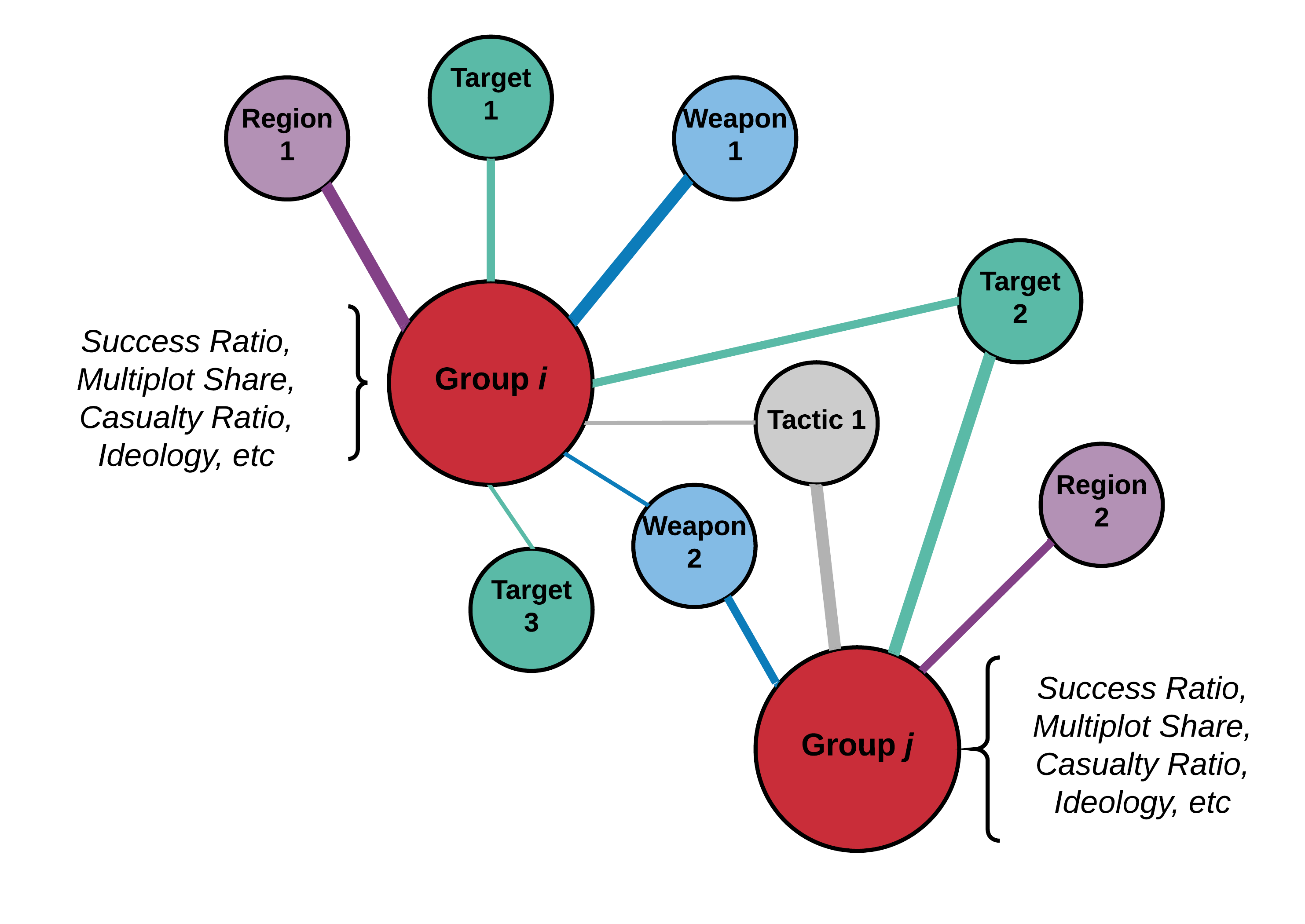}
\caption{\csentence{Example of a Multi-partite Network}
     The graphic illustrations shows a multi-partite network that includes two groups $i$ and $j$}
\label{structure}
\end{figure}

Since the variables of the modes of Targets, Weapons, Tactics, and Regions form a many-to-many relationship with the terrorist groups, we first model this data as a multi-partite network (Figure \ref{structure}) with each partition joined to the terrorist groups; this is often referred to as a `Star' structure with the partitions. 

More specifically, we define: 

\begin{equation}
\mathfrak{G}^N:=\langle\left( V_{1},V_{2},\cdots, V_{n} \right), \left( E_{1,2},E_{1,3}, \cdots,E_{m,n} \right), \left( W_{E1,2}, \cdots,W_{Em,n} \right)\rangle
\end{equation}
as a multi-partite graph that contains $N$ partitions describing relations between different sets of nodes $V_{m}$ and $V_{n}$: these relations are formalized as edges $E_{m,n}$ that are weighted by $W\in \mathbb{\mathbb{R}}_{\geq 0}$ and each mode in the multi-partite network is represented as $G_{m,n}:=\left \langle \left(V_{m},V_{n} \right ), E_{m,n}, W_{E_{m,n}} \right \rangle$. With this data structure we then employ Gower’s Coefficient of Similarity \cite{gower_general_1971} to place the groups in a latent space, whereby we can create a latent network of the groups and assign groups to clusters based upon the multi-partite network. In the latent network, the edges maps the similarity between group $i$ and $j$, calculated using Gower’s Similarity Coefficient defined as: \color{black}
\begin{equation}
S_{ij}=\frac{\sum_{k=1}^{n}w_{ijk}S^{(k)}_{ij}}{\sum_{k=1}^{K}w_{ijk}}\label{sij}
\end{equation}
where $S_{ij}$ is the similarity between terrorist groups $i$ and $j$ on a variable (i.e. Targets, Weapons, etc.), $k$, and $K$ is the total number of variables across all $N$ modes, and $w_{ijk}$ is the weight of the similarity between group $i$ and group $j$ for variable $k$. $S^{(k)}_{ij}$ is then dually defined as: 

\begin{equation}
S^{(k)}_{ij}:\left\{\begin{matrix} 1, & if (x_{ik}=x_{jk}) \neq \emptyset\\ 0, & otherwise \end{matrix}\right.
\label{sijk_dummy}
\end{equation}
if the variable, $k$, is categorical (to include binary) for node $i$ and $j$'s responses, $x_{ik}$, $x_{jk}$, and: 

\begin{equation}
S^{(k)}_{ij}: \frac{\left | x_{ik}-x_{jk} \right |}{r_{k}}\label{sijk}
\end{equation}
where $r_{k}$ is the range of $x_{k}$, if $k$ is numerical. For each variable, $k$ that is numerical the range is calculated as:

\begin{equation}
r^k = |max(x_k) - min(x_k)|
\end{equation}
which means that the range is given by the absolute value of the maximum value of the variable $k$ minus the minimum value of the variable $k$. Gower's coefficient of similarity provides a wide degree of flexibility as it can take various data types, like integer, binary, or continuous values and does so without the use of dummy variables. So, with this coefficient of similarity we are able to incorporate various means of describing terrorist groups, which can be of nearly any data type, and do so in such a way that keeps the original structure of each of the modes as bipartite networks intact. 

Another advantage of Gower's Coefficient is the weighting term. As was noted at the beginning of this section, each of the possible variables used to find similarity between terrorist groups are not independent, but rather fall into various related modes. For example, if group $i$ has operated in the Middle East, it is possible that it has also operated in North Africa or Southwest Asia. Furthermore, since the relationships within the mode are many to many (e.g. a terrorist groups can use many different weapons and vice-versa) each of these modes are a bipartite network. Thus, our data is modeled as a collection of bipartite networks, where there are relationships between the entities within each of the mode networks. So, to take advantage of this model of our data, we employed \textit{network entropy} as the means to weight different modes for the weighting scheme in the Gower's Coefficient \cite{passerini_entropy_2008}. Von Neumann's network entropy is a spectral measure  originating from Gibbs entropy that has been applied to the quantum realm and that provides information on the complexity of a graph and on the amount of information that a network contains. In general, network entropy can be thought of as a measure of how heterogeneous a network is in terms of its connections \cite{passerini_entropy_2008}, \cite{feng_entropy_2019}. As such, network entropy has been used to characterize changes within dynamic graphs, as it is good as distinguishing different graph snapshots from each other \cite{Ye_entropy_2018}. Furthermore, network entropy has also been used a means of distinguishing certain graphs from each other, with those graphs that have a higher entropy having more complex structures like subgroups \cite{passerini_entropy_2008}, \cite{ye_entropy_2014}. So, we similarly employ network entropy to distinguish between the different modes, which are bipartite networks, in such a way that those modes that have more heterogeneous structures --- and as such are likely to be better for separating terrorist groups into clusters --- are considered as more important. Since Gower's Coefficient allows for weighting for exactly the purpose of emphasizing more important features in data, we can use network entropy with the weighting term in Gower's Coefficient to automatically emphasize more useful modes of our data. Following the derivations of the entropy of a network in \cite{passerini_entropy_2008} and \cite{silva_dynamic_2015}, we define the entropies of each of our modes as: \color{black}

\begin{equation}
    H^{n} = -\sum^{|V|}_{i=1}\frac{\Tilde{\lambda_i}}{|V|} ln \frac{\Tilde{\lambda_i}}{|V|}
\end{equation}
where $\Tilde{\lambda_i}$ are the eigenvalues of the normalized Laplacian of the graph of the particular mode. So, for $n \in N$ the normalized Laplacian is $\Tilde{L}^{n} = D^{-\frac{1}{2}}(D-X^{n})D^{-\frac{1}{2}}$, where $(D-X^{n})$ is the unnormalized Laplacian and therefore $X^n$ is the adjacency matrix from a particular mode and $D$ is the degree matrix, which is created by:

\begin{equation}
D:\left\{\begin{matrix} \sum_{j=1}^V X_{ij} & \mathrm{if}\; D_{ii} \\ 0 & \mathrm{otherwise} \\ \end{matrix}\right.
\end{equation}
Following the findings on using network entropies to characterize heterogeneous graphs in \cite{Ye_entropy_2018} and \cite{feng_entropy_2019}, we let those modes with higher entropy have more impact on the similarity measurements through Gower's coefficient. So, the weighting term in our Gower's coefficient is:

\begin{equation}
    w_{ijk} = \sum_n^N \delta(k,n) \times H^{n} 
\end{equation}
where $\delta(k,n)$ is an indicator function that returns 1 if variable $k$ is in mode $n$, and 0 otherwise. It should be noted that each variable within a mode will recieve the same weight. So, those modes which have a more heterogeneous structure, which should be better for producing structures like clusters, will have a higher weight in the comparison of the various terrorist groups. \color{black}

\subsection*{Asymmetric kNN Modularity Graph Construction}
Having obtained pairwise similarities between all of the terrorist actors, we now move on to extracting a network from the data, which we refer to as the latent network. Following our work in \cite{CampedelliDetectingLatentTerrorist2019}, we continue to use the kNN modularity maximization procedure proposed in \cite{ruan_fully_2009}. At a a high level, once similarities have been computed for each of the terrorist groups, \color{black} the method iterates through various possible numbers of neighbors for each node, $k$, and selects that $k$ which produces the most modular graph, relative to a null-model, random graph produced on the same similarities. \color{black} Modularity in this case is the network modularity as described in \cite{Newman_networks_2010}:

\begin{equation}
    mod(G) = \frac{1}{2m}\sum_{ij}[ A_{ij} - \frac{deg(i)\times deg(j)}{2m} \delta (c_i, c_j) ]
\end{equation}
where $m$ is the number of links in the network, and $c$ are the cluster assignments of the nodes. A graph with high modularity is one which will have sub groups that have a lot of interconnections. Since it is known that random graphs can give rise to modular structures, we also compare this modularity value to the modularity value obtained from a random graph with the same number of vertices and edges, and the same similarities between the vertices \cite{ruan_fully_2009}. The general idea behind this method is that a kNN that is higher in modularity, relative to a null-model, is better for detecting community structure in the underlying data used to make the network. It should be noted that this procedure applies only after a measure of similarity has been applied to the data.  \color{black}
\\
We have, however, modified the algorithm slightly to better suit our data. First, we use an asymetric kNN network. More precisely, for each point $i$, let $N_k(i)$ be the $k$ nearest neighbors of $i$, then an asymmetric kNN network has links between two nodes $i$ and $j$ if $i \in N_k(j)$ OR $j \in N_k(i)$. Second, in the clustering step, we differ from the original algorithm proposed in \cite{ruan_fully_2009}, as we use a faster method of modularity maximization of unimodal networks, the Louvain Method \cite{blondel_fast_2008}, as opposed to the author's \textit{QCut} algorithm. The psuedo-code of our implementation of network construction by kNN modularity maximization is detailed in algorithm \ref{kNN}.
\\
\begin{algorithm}
\caption{kNN Modularity Maximization Procedure}
\label{kNN}
\begin{algorithmic}
\BState \textbf{input}: Distance or Affinity Matrix, $S$, (n x n)
\BState \textbf{output}: Optimal k-Nearest Neighbor Network, $G^*$, and sub group assignments $C(G^*)$.
\For{$i=1 : \left \lfloor{log_2 (n)}\right \rfloor $}
	\State $k \gets 2^i$
	\State $G_k \gets kNN(S,k)$
	\State $C(G_k) \gets Louvain(G_k)$
	\State $G^r_k \gets randomize(G_k)$
	\State $C(G^r_k) \gets Louvain(G^r_k)$
	\State $Modularity_k \gets Modularity(C(G_k)) - Modularity(C(G^r_k))$
\EndFor 
\State $k^* \gets argmax_k Modularity_k$
\State $G^* \gets kNN(s, k^*)$
\State $C(G^*) \gets Louvain(G^*)$ \\
\Return $C(G^*)$, $G^*$
\end{algorithmic}
\end{algorithm}

In Algorithm \ref{kNN}, $G_k$ is a particular kNN graph, where each vertex connects to exactly k of its nearest neighbors. The sub-step of $randomize(G_k)$ is to randomly re-wire all of the edges in $G_k$. This is equivalent to creating an Erdos-Renyi random graph that has the same number of edges and vertices as $G_k$. This step is performed in order to create a null-model of $G_k$, so that we can get a better idea of the strength of the modularity of the proposed $G_k$ by comparing it to the modularity of its null counterpart, $G^r_k$. So, a good kNN should not just have high modularity, but also high modularity with respect to a randomized version of that kNN; the modular structures should not be just an artifact of the kNN's density or size. Finally, we return $G^*$ which is that kNN which has the most modular structure. \color{black} A Python implementation of the code will be available on the author's GitHub page with publication of this article. 

\section*{Results}
\subsection*{Comparing clustering assignments}
\begin{figure}[h!]
\includegraphics[scale=0.28]{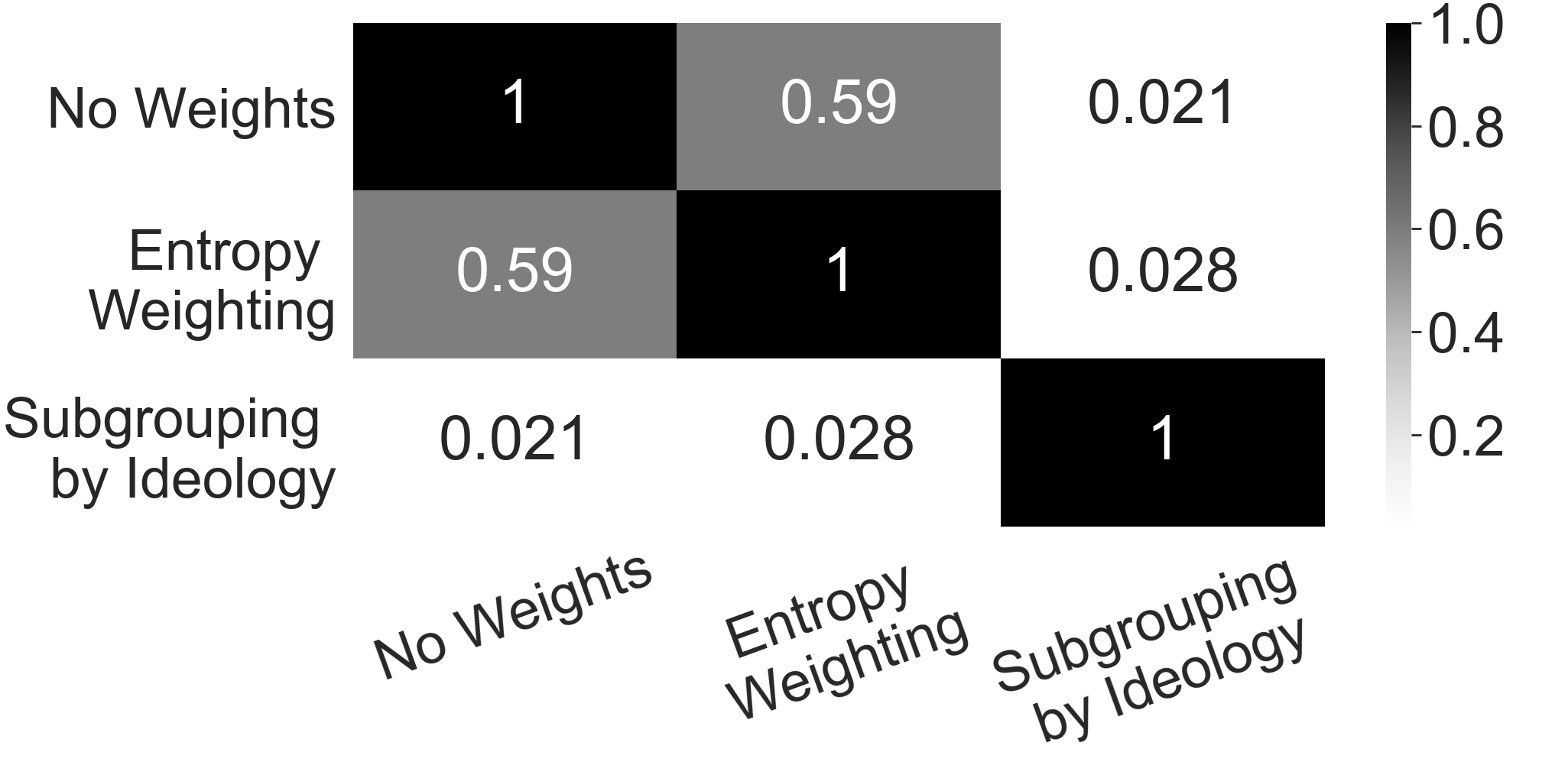}
\caption{\csentence{Adjusted Mutual Information of three Subrgrouping Schemes.}
     The ideology-based heuristic approach produces subgroupings which are extremely different when compared with the baseline and the entropy-based approach}
\label{ami}
\end{figure}

To compare the similarity of clustering assignments of the three grouping procedures we have calculated the Adjusted Mutual Information (AMI), which is a modified version of the ordinary mutual information adjusted for randomness, and it is calculated as
\begin{equation}
    AMI(U,V)=\frac{MI(U,V)-E\left \{ MI(U,V) \right \}}{max\left \{ H(U), H(V) \right \}-E\left \{ MI(U,V) \right \}}
\end{equation}
where $E\left \{ MI(U,V) \right \}$ is the expected mutual information between two random clusterings and $H(U)$ and $H(V)$ are the entropies associated to each partition $U$ and $V$. This is a standard way to measure how different or similar are the outcomes of clustering procedures. The results are displayed in Figure \ref{ami}.

The results clearly show how, with respect to the baseline unweighted model, subgrouping using entropy weighting is far more similar than the ideology-based subgrouping. This, on one hand, suggests that relying only on this latter heuristic can extremely underestimate and distort the latent similarity that exist across groups when fully considering behavioral or operational variables. While ideology is certainly important for contextualizing a certain terrorist actor, the multi-partite original network includes information that are not captured by this method. 

On the other hand, the entropy-based weighting is able to capture a relatively high portion of the information associated with the unweighted baseline model. While certainly introducing this data-driven discriminatory procedure affects the final clusters, our first conclusion is that this method is far more reliable if we want to preserve the original information structure of our data. Moreover, this algorithmic approach might provide a more solid tool to analysts and policymakers if they need to go beyond the original data, exploiting the richness of the original data itself. 

Given these results, we now proceed to compare the baseline and the entropy-based clusterings more in depth, to understand whether, besides pure group assignments, they also share stable similarities in terms of behavioral and ideological features.

\subsection*{Unweighted vs Entropy-based Subgrouping: Similarities and Differences}

Our algorithmic procedure to create the k-Nearest Neighbor network yielded two different graphs. Focusing on global characteristics of both networks, we can highlight how these graphs hold distinct structural and topological characteristics (Table \ref{global}). 
\begin{table}[!h]
\caption{Network structural and topological characteristics for both approaches}
\label{global}
\begin{tabular}{lcc}
\hline
                            & \multicolumn{1}{l}{\textbf{Unweighted}} & \multicolumn{1}{l}{\textbf{Entropy-based}} \\ \hline\hline
Best k                      & 2                                       & 4                                          \\ \hline
Modularity                  & 0.805                                   & 0.390                                       \\ \hline
N of Links                  & 5,242                                   & 10,658                                     \\ \hline
Bi-directional Link Count   & 2,621                                   & 5,329                                      \\ \hline
Density                     & 0.002                                   & 0.004                                      \\ \hline
Clustering Coefficient      & 0.342                                   & 0.336                                      \\ \hline
Betweenness Centralization  & 0.656                                   & 0.596                                      \\ \hline
Eigenvector Centralization  & 0.826                                   & 0.703                                      \\ \hline
Total Degree Centralization & 0.106                                   & 0.257                                      \\ \hline
\end{tabular}
\end{table}

The table above highlights how the $k$-NN procedure provided two different optimal k's for the two networks. The entropy-based network has a higher k, and this justifies the higher number of links (both overall and bi-directional), and the higher network density. Nonetheless, the unweighted network proves to be higher in clustering coeficient, betweenness centralization and eigenvector centralization, while the entropy-based one yielded higher total degree centralization. With regard to modularity, the unweighted network notably performs a higher value, suggesting that the emerging clusters are more defined than the ones yielded by the entropy-based approach.

\begin{figure}[h!]
\includegraphics[scale=0.55]{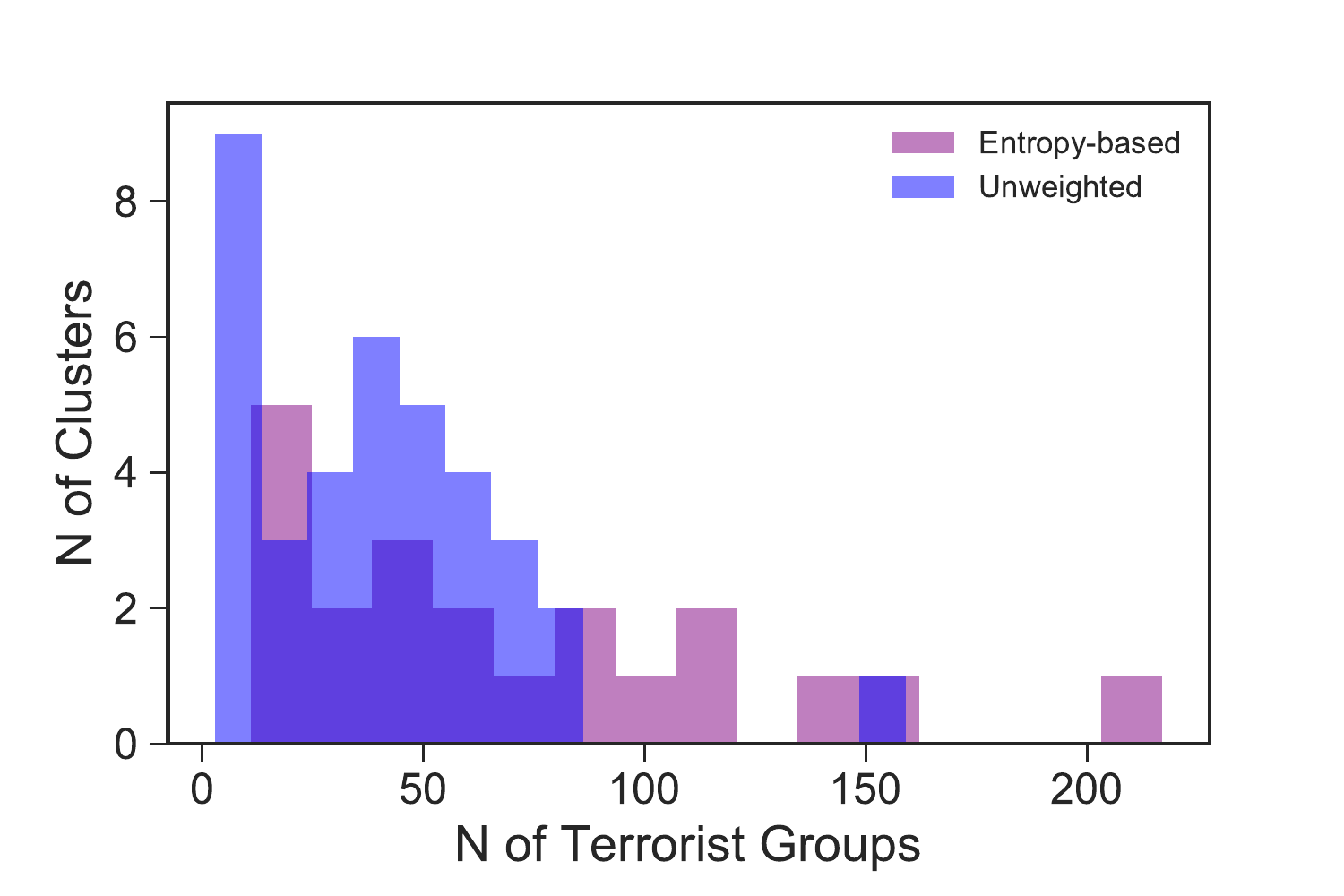}
\caption{\csentence{Distribution of Cluster Assignments per Different Approach}
      The unweighted approach produces a higher number of small clusters, highlighting a higher modularity.
      \label{hist}}
      \end{figure}

Connected to this aspect, and with regard to the actual subgroupings, is the fact that the entropy-based procedure produced less clusters (21) compared to the unweighted one (37). As Figure \ref{hist} shows, the entropy-based approach produces a greater number of highly populated clusters, while in the unweighted case, a considerable amount of clusters includes a little number of groups (in fact, 25 clusters include less than 50 terrorist groups each). These figures further justify the different scores in term of modularity, since a higher number of smaller clusters is highly likely to indicate a higher degree of diversity in the network itself, as captured by modularity. 

\begin{figure}[h!]
    \centering
    \includegraphics[width=.45\textwidth]{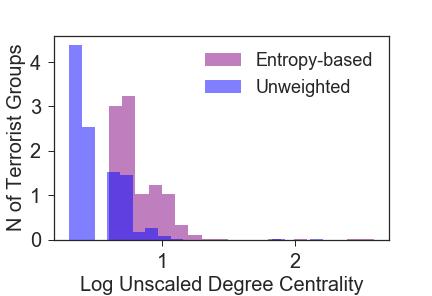}\quad
    \includegraphics[width=.45\textwidth]{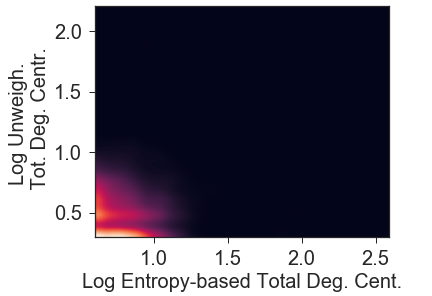}
    
    \medskip
    
    \includegraphics[width=.45\textwidth]{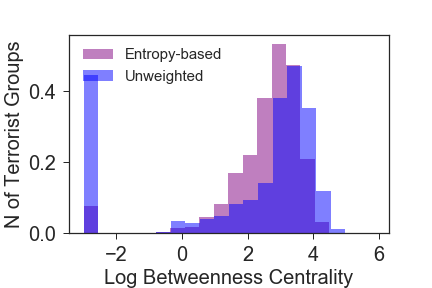}\quad
    \includegraphics[width=.45\textwidth]{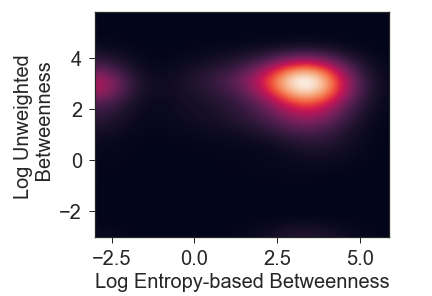}
    
    \medskip
    
    \includegraphics[width=.45\textwidth]{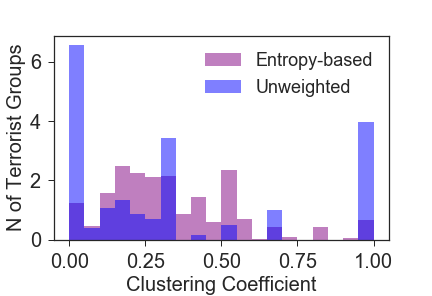}\quad
    \includegraphics[width=.45\textwidth]{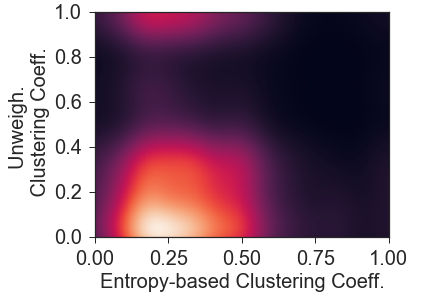}
    
    \caption{\csentence{Histograms (20 bins) and 2D Kernel Density Estimation of Log Unscaled Total Degree Centrality, Log Betweenness Centrality and Clustering Coefficient of Terrorist Groups across both approaches}
    2D Kernel Density Estimations are performed through regular grids made of 500 bins for each axis.
    \label{fig:node_level}}
\end{figure}

Focusing on the different node-level measure distributions, Figure \ref{fig:node_level} displays histograms and 2D Kernel Density Estimations (KDE) of three selected metrics, namely Log Unscaled Total Degree Centrality, Log Betweenness Centrality and Clustering Coefficient. Total Degree Centrality and Betweenness Centrality have been transformed in log scale in order to provide more intuitive graphic results, since the original distributions are extremely left-skewed and the bivariate visualizations would have been extremely difficult to interpret. \color{black}
\\

\begin{figure}[h!]
\includegraphics[scale=0.3]{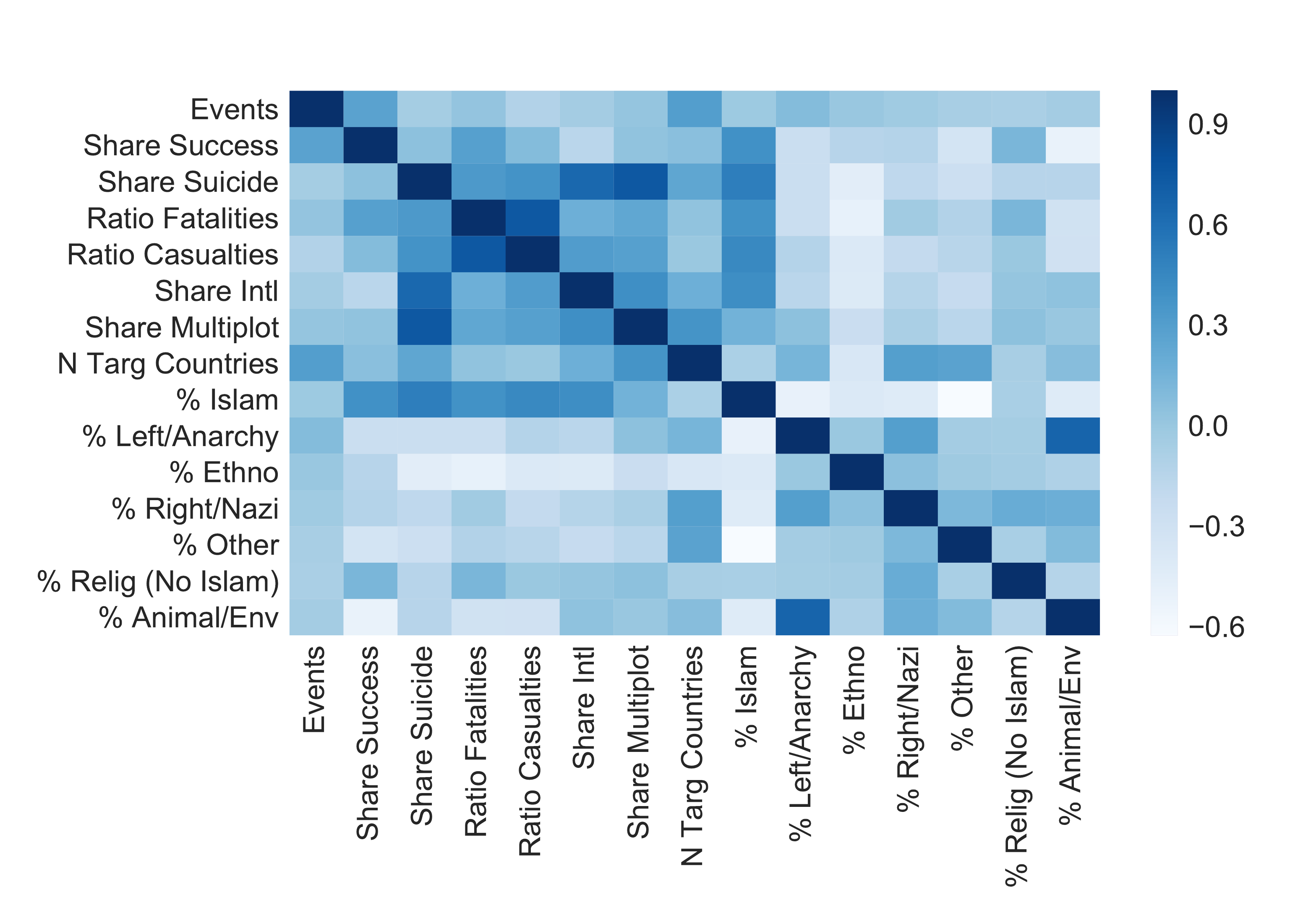}
\includegraphics[scale=0.3]{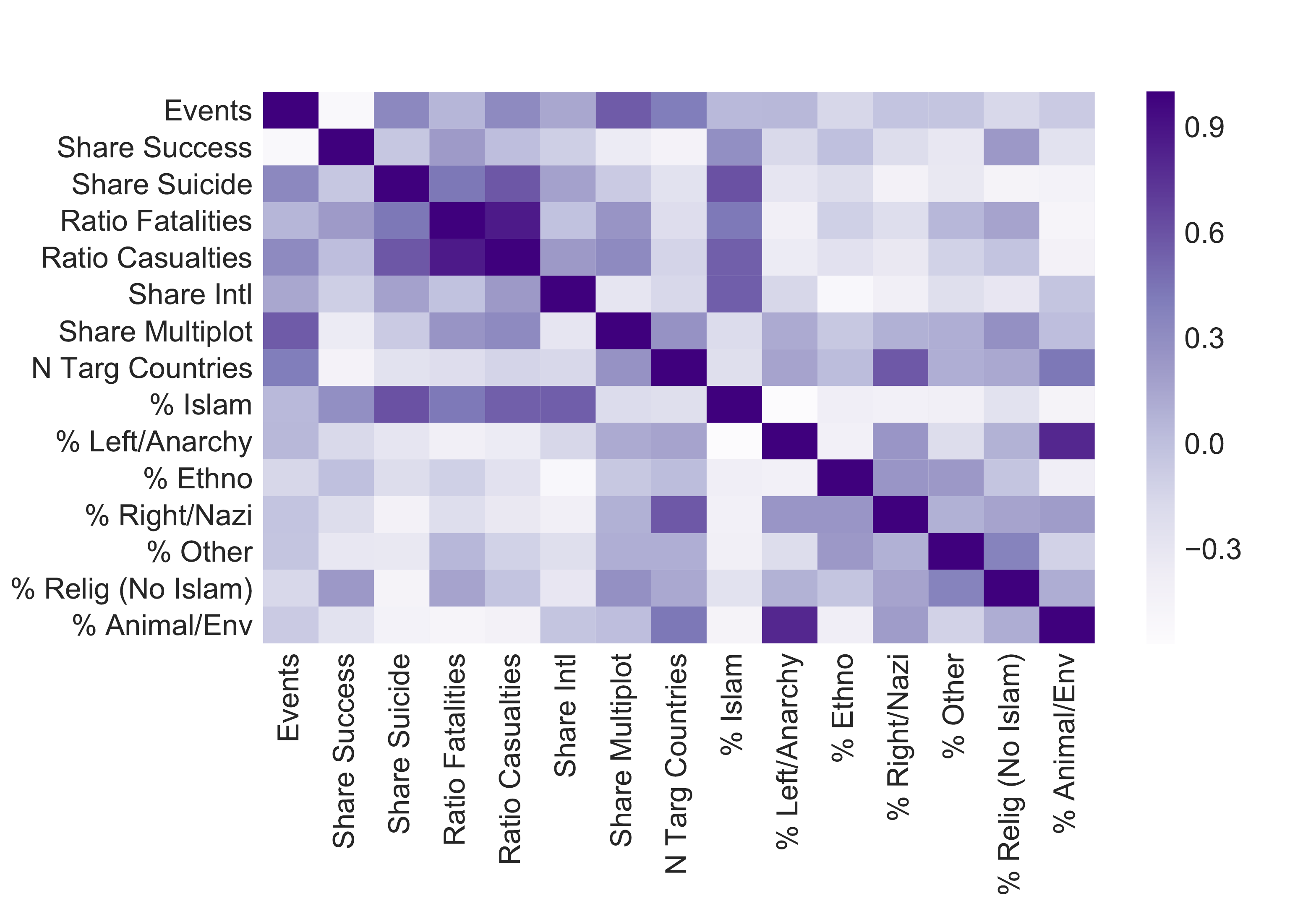}\quad
\includegraphics[scale=0.28]{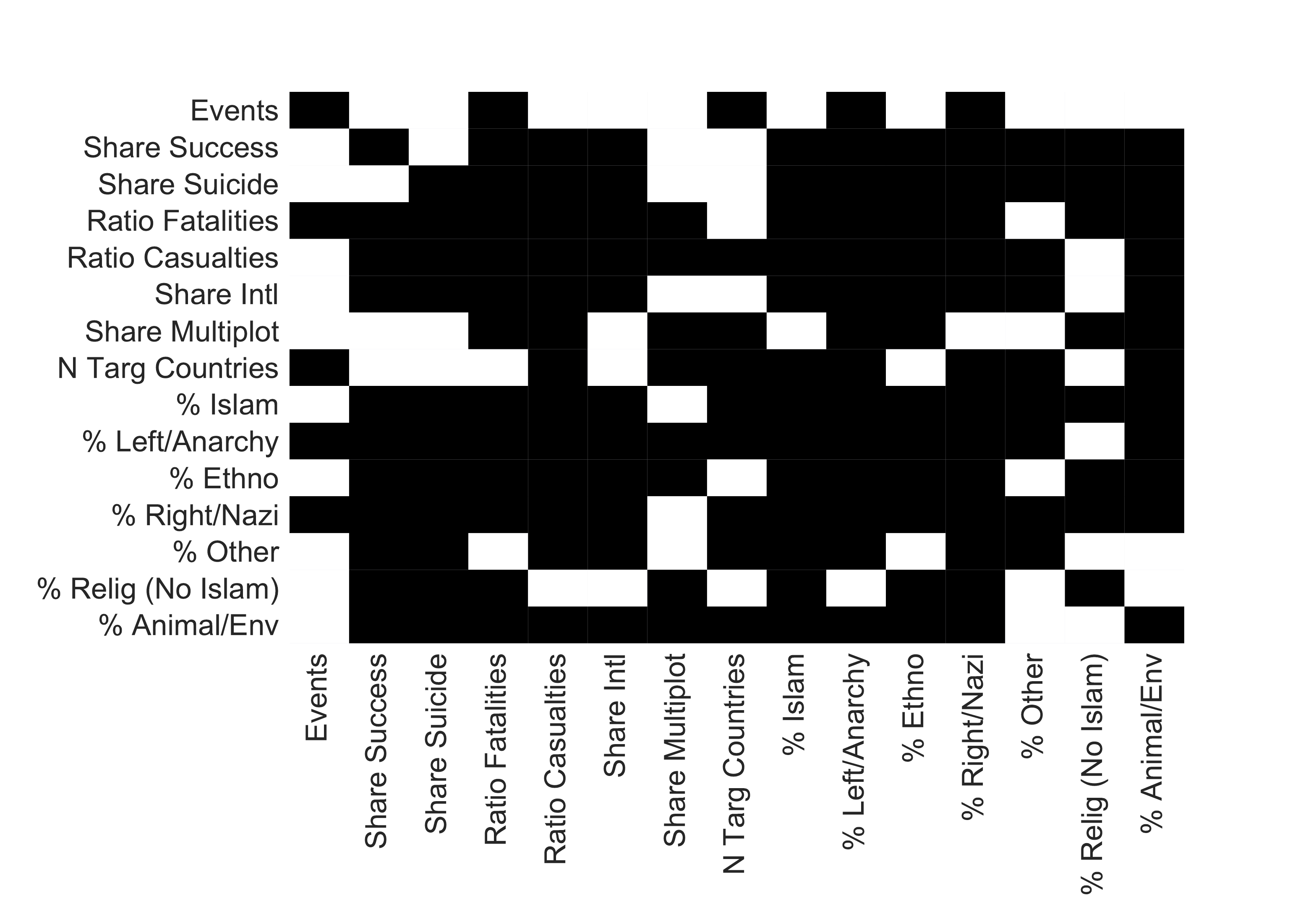}
\caption{\csentence{Correlation among behavioral and ideological variables Unweighted (left) and Entropy-based (right) approaches}
      Correlation shows what bivariate relations stay stable across both approaches (namely, they maintain the same direction). Bottom plot highlights (in black) stable relations against non-stable (white).
      \label{corr_comp}}
      \end{figure}

Regarding Log Unscaled Total Degree Centrality, the histogram highlights how the groups in the Entropy-based approach generally have less connections that the ones in the Unweighted approach. The 2D KDE displays a strong concentration of data points in the bottom-left side of the graph, with more dense concentrations. 
\\
With regards to Log Betweenness Centrality, the histogram shows relatively similar distributions for the considered approaches. The Unweighted one displays a greater number of  groups with betweenness equal to zero (log($n)\approx-3).$\footnote{When betweenness was equal to 0 we have transformed it to 0.001 in order to allow the value for log transformation. This transformation did not affect the results since it was performed only to provide intuitive visualizations and interpretable bivariate relations} \color{black} The 2D KDE displays a very high concentration of nodes on the top right of the plot, showing a positive correlation of log betweenness centrality across nodes for both approaches. However, it is worth noting that there is also an interesting small concentration of nodes that have very high value in the Entropy-based case but, conversely, very low ones in the Unweighted case.
\\
Conversely, in relation to clustering coefficient, more evident differences emerge when looking at the histogram. In fact, the Entropy-based approach shows a more concentrated distribution, while the Unweighted one highlights a very different behavior, with a considerably high number of extreme values, on both left and right side of the x-axis. However, these differences are mitigated in the KDE plot. Indeed, it shows the concentration of the majority of data points in the bottom left, almost indicating a linear relationship. In spite of this, it is wort noting that there exist a portion of groups which obtain very high levels of clustering coefficient in the Unweighted case, while their corresponding values in the Entropy-based approach are significantly lower. In light of the considerations on these detected differences in topology, structural and node-level measures of both networks and cluster formation, it is worth inspecting the types of groups in terms of operational and behavioral features and ideologies that are clustered together in both approaches.\\
The correlation results interestingly showcase that the majority of relations (either positive or negative) hold stably across both approaches, while only few have opposite directions from one approach to the other (Figure \ref{corr_comp}). Notably, correlations on events (first columns of both plots) are generally very different. This may suggest that the raw number of events do not drive any consistent information flow regarding cluster assignments. This would indicate that there are other types of features that actually capture similarities or differences across terror groups, and that the latent data structure is independent from the individual frequency of attacks of actors. In terms of stable results, both approaches demonstrate how clusters with a high percentage of islamist or jihadist groups are associated with high levels of attack success, while this type of relation goes in the opposite direction for all the other ideologies. This marks a distinctive feature of jihadism or islamism as terror motivation. Expectedly, suicide attacks are also found to be positively correlated with jihadist or islamist ideology. Furthermore, Islamist/jihadist ideology is again the only one positively associated with high levels of both casualties and fatalities in both approaches, while other ideologies seem to be less lethal. In terms of the multiplot, which captures the extent to which a terrorist group is able to plot multiple coordinated attacks in the same day as part of a more complex logistic structure, FL groups, along with religious (non-Islamist) groups, are the only ideologies to display a stable, positive relation. 

\begin{figure}[!h]
\includegraphics[scale=0.39]{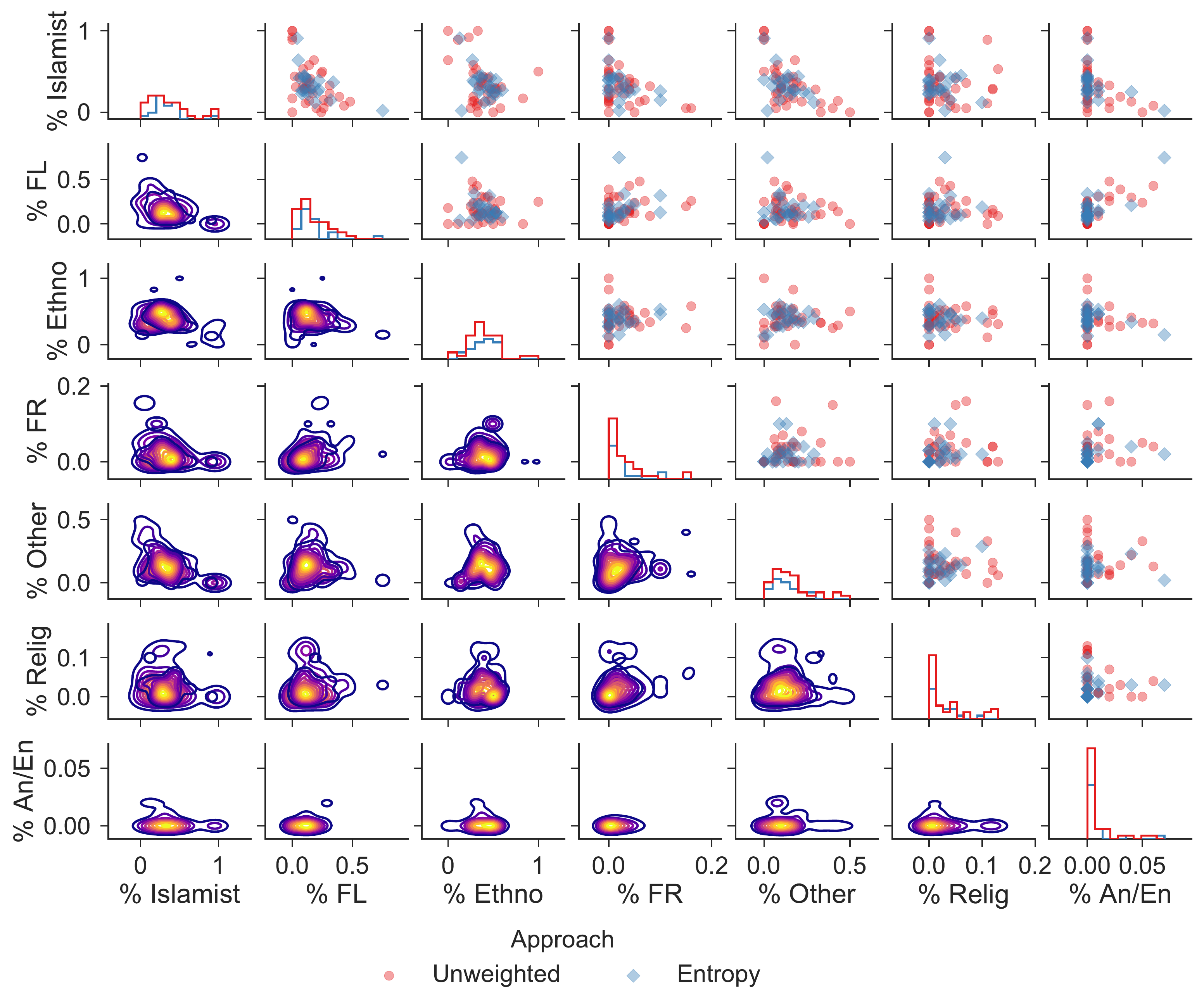}
\caption{\csentence{Kernel Density, Histogram and Scatter Plots of Cluster Distribution for Both Approaches - Ideological Variables}
The graphic visualizations demonstrates that more represented or well-defined ideologies tend to have stabler relations across both approaches. Relation for environmental/animalist and religious (Islamist excluded) groups appear to be more volatile.
      \label{corr_entro}}
\end{figure}

Clear results emerge when focusing on pairwise relations between ideologies (Figure \ref{corr_entro}). Overall, the majority of relations are stable across both algorithmic approaches. Islamism has negative correlations in both cases with all the other ideologies, implying that groups belonging to or motivated by this ideology represent very distinguished entities in the global terrorist scenario. With regard to FL groups, they strongly share similar cluster assignments with enviromentalist and animalist groups and, surprisingly, they share similar assignments also with FR actors.
\\
This result suggest that, while these two ideologies are considered very different from one another and the motivations of groups belonging to these factions are extremely distant, from the operational point of view (namely, from the standpoint of employed weapons, hit targets, applied tactics and targeted regions) FL and FR terrorist groups are quite similar, and this result is corroborated by its stability across the two approaches. FL groups are not the only actors that share cluster assignments with FR terrorists. In fact, the analysis show that the higher the fraction of FR groups, the higher the number of religious (Islam excluded) and ethno/nationalist actors. 

These type of relations might be expected, considering that many FR groups include elements of bigotry or radical religious views and that many nationalist actors generally rely upon fascist or far-right discourse and political behaviours. However, another surprising result is given by the negative relation between FL and ethno/nationalist groups, considering two factors: first, the positive stable correlation between FR and FL groups; second, the fact that radical leftist or communist ideologies are generally the other opposite driving force of certain nationalist or independence-driven actors, such as the Euskadi Ta Askatasuna (ETA) in Spain. This means that, besides the potentially similar motivations and background, these two types of groups act and organize attacks that are generally dissimilar. 

\subsection*{Potential Directions: Introducing a Dynamic Entropy-based Approach}
\begin{figure}[h!]
    \centering
    \includegraphics[width=.65\textwidth]{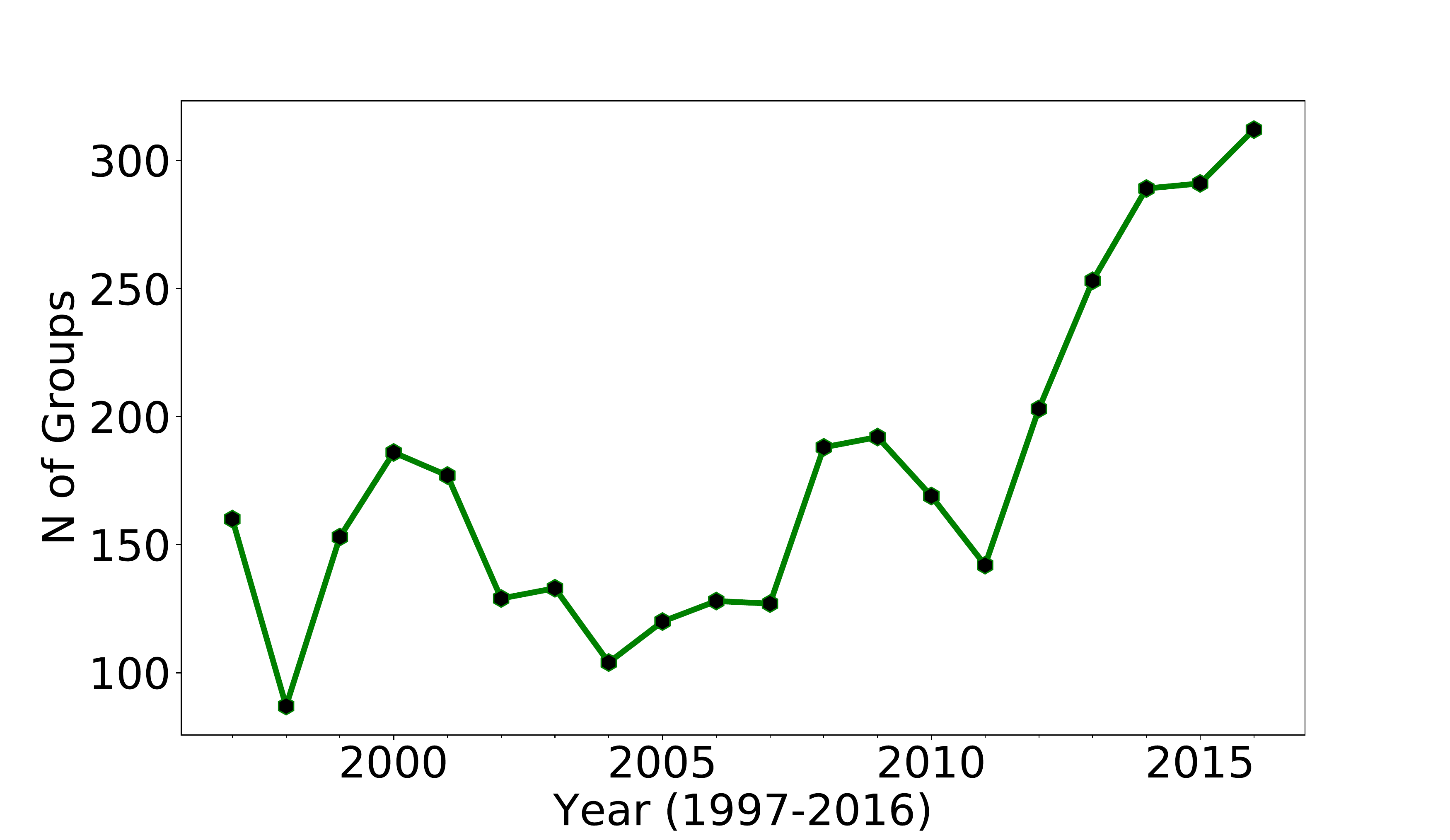}
    \includegraphics[width=.65\textwidth]{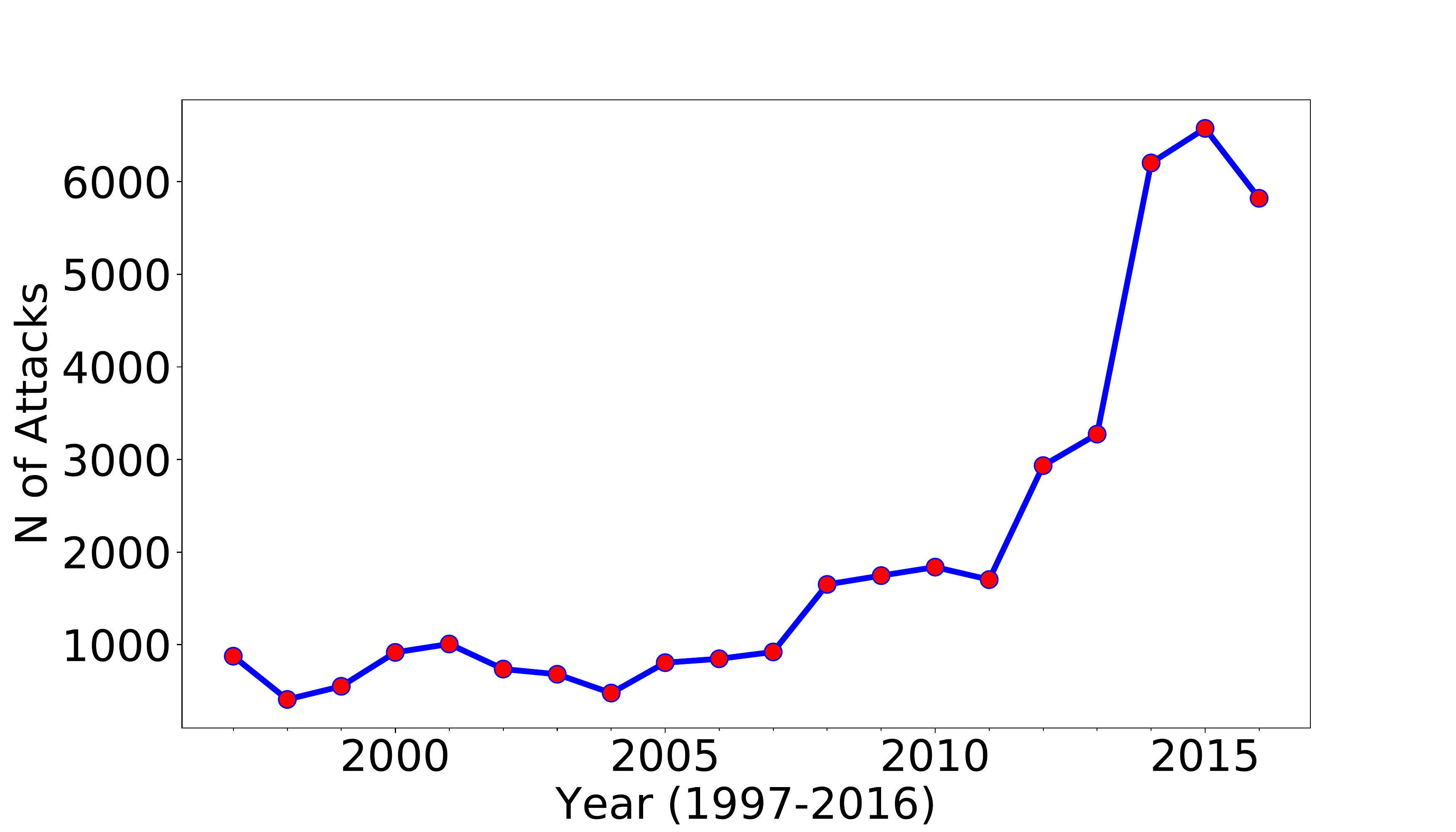}
\caption{\csentence{Number of Active Groups and Terror Attacks per Year (1997-2016).}
    \label{active}}
\end{figure}

The comparative analysis across the two approaches demonstrated that the entropy-based approach preserves a relatively high amount of outcomes derived from the baseline model, also shedding light on additional mechanisms when focusing on the $ex$-post analysis of behavioral and ideological features. However, our experiment comes with a limitation that should be adressed in the future. In fact, our original multi-partite network spans across twenty years and includes the cross-sectional information on the whole universe of active groups without taking into account time in a dynamic fashion. 
\\
Terrorism has undergone several changes in the last two decades: many new groups have appeared only recently, many have disappeared or have been dismantled, other actors have been active only for very short period. In general, over the years, the trend of active actors has not shown a stable behavior. As it is expected, this type of trend also regards the number of attacks and terrorist events (Figure \ref{active}).

\begin{figure}[h!]
\includegraphics[scale=0.45]{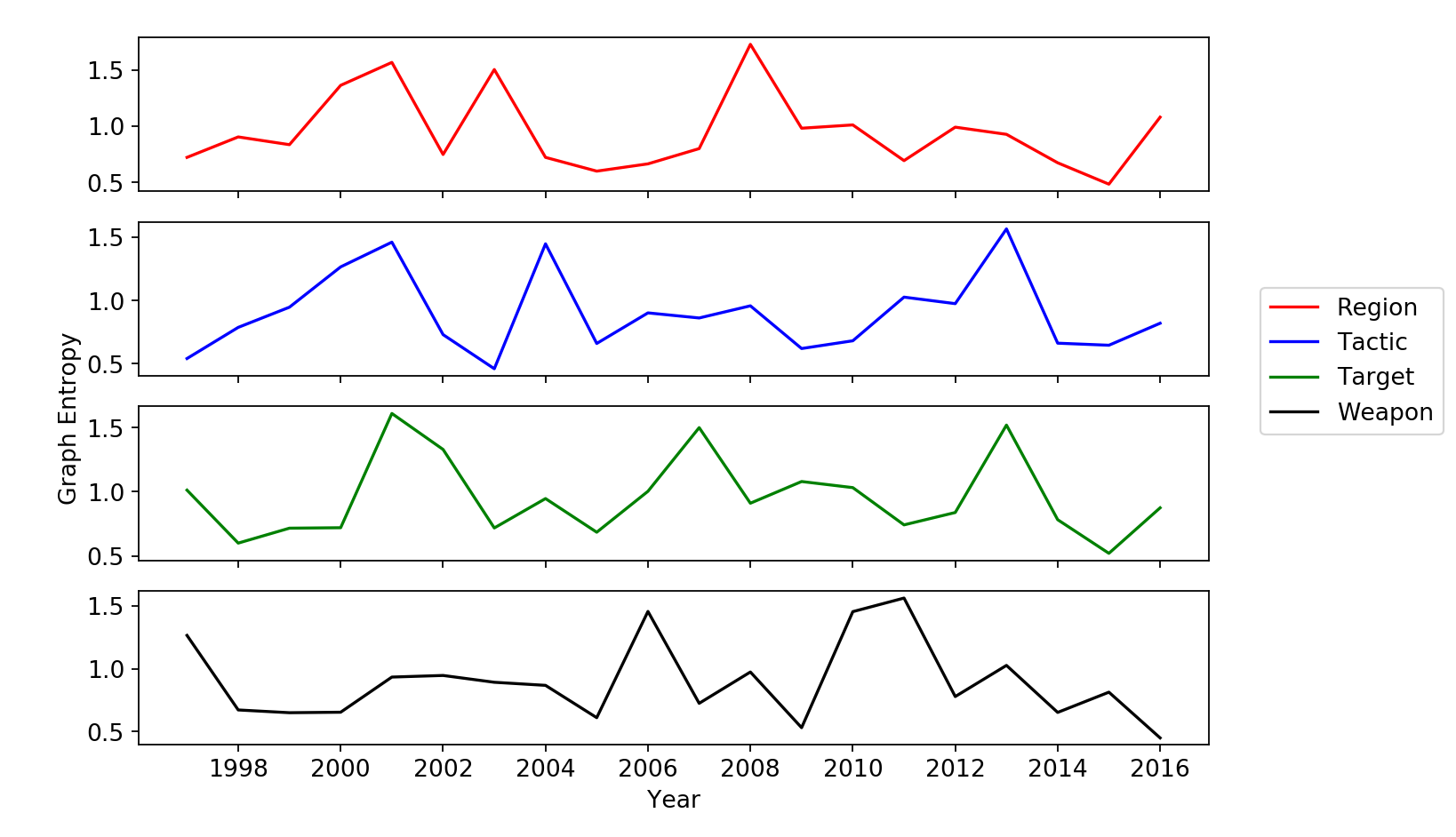}
\caption{\csentence{Entropy variation for each mode (1997-2016).}
      \label{overtime}}
      \end{figure}

These changes and trends over the last twenty years may also be related to strategies and types of attack, besides mere frequency of attacks. This is an aspect which intersects our analysis. Since our original multi-partite network stores all the information of the past twenty years, these changes and trends may be underestimated or even vanish altogether in our algorithmic procedure. As for the current data structure, groups that have plotted very few attacks in two distant years may be clustered together considering their similarity in the data space, however, it might be not useful for analysts or policy-makers to compare two groups that are too distant in time. For this reason, our intuition is to introduce a potential solution to this issue via a dynamic Entropy-based approach. Instead of constructing a single static multi-partite network, we build yearly multi-partite graphs to capture the variations in the entropy of each mode (Figure \ref{overtime}). 

As shown by the plot above, there exists clear variations in the entropies of each mode from which we have built our multi-partite network. The plot highlights how the Weapon mode generally follows completely different trends with respect to all the others. At the same time, the entropies of Tactic and Target modes display very similar behavior in the last four years of our considered timespan. Notably, there is an almost complete similarity of entropies for Region, Tactic and Target modes with respect to year 2001. This graph clearly suggests that time should be considered in the algorithmic procedure, since it is highly likely that sensible changes in the subgrouping outcomes will emerge. Besides the relevance of embedding temporal dynamics from the purely research point of view, this furthermore provides a richer tool for potential users and analysts interested in using the algorithm to detect, assess and study patterns in the data. In fact, since our model seeks to provide a practical framework that can be easily deployed for the aforementioned purposes, we feel that time-aware results are able to exclude all the non-active groups in a given year and would increase the usability of previous years information and its efficiency for real-time objectives oriented to intelligence profiling. 

\section*{Discussion \& Future Work}
This work has presented a novel algorithmic framework for detecting latent clusters of similar terrorist groups via a complex network approach. We have created a multi-partite network for the entire known population of terror groups active worldwide from 1997 to 2016, where modes were Weapons, Tactics, Targets, operating Regions, and proposed a novel clustering architecture expanded from \cite{CampedelliDetectingLatentTerrorist2019}. We have then compared our new entropy-based architecture with two alternative solutions: a weak-heuristic approach based on terror group clustering by ideology and our baseline unweighted approach. The entropy-based approach modifies the baseline approach simply weighting each mode by its graph entropy, in order to provide a data-driven approach that takes into account the relevance of a certain mode with respect to the others. The analysis has first demonstrated that subgrouping by ideology leads to cluster assignments very different to the ones obtained with our baseline method, where we let patterns emerge naturally from data with no \textit{a priori} knowledge and, secondly, that the entropy-based and the baseline approaches have similar results both in terms of stability of cluster assignment for terrorist groups and behavioral and ideological intra-cluster association.   
\\
Both approaches corroborated interesting findings that go beyond the pure methodological intent of this work. To investigate the meta-connections between groups resulting from our work, we have analyzed behavioral characteristics (e.g., share of successful attacks, international propensity, etc.) and we have also focused on the ideological background of each actor, retrieving this information from BAAD version 1 and 2 and other open access qualitative sources. Though labelling a group under few ideological categories may oversimplify certain complex components of terrorism, interesting relations emerged. \color{black} Besides several expected patterns (e.g. Islamist/jihadist groups tend not to be associated with groups belonging to other ideologies), the algorithm reveals other results that may shed light on terrorism in terms of research and policy. The clustering procedures highlighted a certain similarity between FL groups and FR groups, indicating that besides their divergent objectives and goals, these two types of groups share similar behaviors. Furthermore, FL groups on one side are often associated with animal-rights and environmentalist actors, suggesting that some overlapping in terms of motives and aims is also connected to similar methods and ways of acting. On the other side, FR groups tend to be clustered together with ethno/nationalist and religious groups, as it might be expected given that many FR groups hold nationalist or religion-related elements.\\ Overall, the entropy-based approach is a flexible tool for capturing the intrinsic and hidden knowledge included in the manifold via a data-driven procedure, rather than using subjective knowledge and weaker heuristics, which was one of the limitations of several experiments conducted in \cite{CampedelliDetectingLatentTerrorist2019}.
\\
In spite of the aforementioned results, our approach may suffer from the fact that the original multi-partite network does not take time into account. Indeed, the manifold includes the whole set of available data from 1997 and 2016. This might be interpreted as a limitation, especially considering that our work is inherently policy-oriented. The last twenty years have been susceptible of several dramatic changes in the ways terrorism manifests itself at global scale. On one side, they have seen the rise of islamist and jihadist terrorism not only in Africa and Middle East, but also in Western and Eastern Europe and countries of the North America. On the other side, politically motivated terrorism has showed shifts and different concentration over time and space. Furthermore, the considered time span is relatively long and therefore includes groups that may have been already disappeared and dismantled or even actors that have plotted one or very few single attacks, therefore constituting  a sort of "noise" in the whole scenario. 
\\
In light of this, we have opened the path for future work showing that, besides variations in the trends of active groups and actors and plotted attacks, there exist also significant variations in the entropies of each mode over time. Entropies change sensibly over-time and we have highlighted the presence of some similarities in these trends across certain modes (e.g. Tactic and Target), while others follow completely different behaviors (e.g. Weapon). For these reasons, future work should test the entropy-based setup within a dynamic framework. While considering all the groups and being able of compare still active groups with those that have been already dismantled or disrupted is certainly useful, we feel that controlling for the noise in the manifold and including only groups that are still part of the global terrorist scenario will provide more insights and will help policy-makers or analysts in understanding to what extent certain groups are similar or different compared to others. Additionally, leveraging upon the entropy-based structure automatically allows one to take into account the most relevant sources of information in real-time (e.g. modes): this type of setup, for instance, would be capable of highlighting anomalous behaviours or strategical behavioral evolution of certain terror groups. Future work will also seek to eventually exclude groups that are not strictly considerable as terrorists (e.g., Mexican drug cartels) although their actions are of terrorist nature: this operation would reduce potential noise and distortion of the results.
\color{black}
\\
Finally, inherent limitations come from the data. While the GTD is certainly the most reliable and solid open access dataset freely available for research-purposes on terrorist events, its structure poses issues of missing data and level of detail of the information. Despite the fact that, as opposed to other criminal phenomena, terrorist attacks tracked and recorded do not face the risk of underestimation (generally, every terrorist attack is reported by newspapers or media agencies), not all details on the attacks might be consistently retrieved and included in the dataset. This would therefore lead to a certain degree of bias or missing information regarding event characteristics, which are the core of our work. Another linked type of limitation is the risk of too generic information, especially for terror attacks occurred outside Europe and North America (which are actually the majority of the events). While our algorithmic framework has demonstrated a certain degree of potential using the GTD, the intent is to test it on more detailed databases in the future. Additionally, in our algorithmic framework, we do not consider any correlation between the different modes of the data. More specifically, it is possible that certain groups use certain weapons or tactics because of limited availability of alternative means and not, instead, as the product of a free choice. Unfortunately, we are not able to assess whether this is the case for the groups under analysis, but this potential explanation shall be kept in mind. Additionally, the ideology labelling process, though based on a scientifically recognized dataset, may oversimplify certain characteristics and motivations behind each group's actions. Reducing the complexity of the causes and motives behind the decision to resort to terrorism is challenging and attention should be payed not to provide distorted or biased interpretation of the results. 
\color{black}


\begin{backmatter}

\section*{List of Abbreviations}
GTD: Global Terrorism Database; BAAD1 and BAAD2: Bad, Allied and Dangerous Dataset versions 1 and 2; kNN: k-Nearest Neighbor; FL: Far Left/Communist/Anarchist; FR: Far Right/Racist/Nazi; AMI: Adjusted Mutual Information; KDE: Kernel Density Estimation

\section*{Declarations}
\section*{Availability of Data and Material}
The datasets analysed during the current study are available in the START repository. The Global Terrorism Database is available at: \hyperlink{https://www.start.umd.edu/gtd/}{https://www.start.umd.edu/gtd/} \cite{national_consortium_for_the_study_of_terrorism_and_responses_to_terrorism_global_2016}; Bad Allied and Dangerous dataset version 1 is available at \hyperlink{https://dataverse.harvard.edu/dataset.xhtml?persistentId=hdl\%3A1902.1/16062}{https://dataverse.harvard.edu/dataset.xhtml?persistentId=hdl\%3A1902.1/16062} and Bad, Allied and Dangerous dataset version 2 is available for browsing at \hyperlink{https://www.start.umd.edu/baad/database}{https://www.start.umd.edu/baad/database} \cite{AsalBigAlliedDangerous2015}.

\section*{Competing Interests}
The authors declare that they have no competing interests.

\section*{Ethics Approval and Consent to Participate}
Not applicable.

\section*{Funding}
This work is supported in part by the Office of Naval Research under  the Multidisciplinary University Research Initiatives (MURI) Program award number N000141712675,  Near Real Time Assessment of Emergent Complex Systems of Confederates, the Minerva program under grant number N000141512797,  Dynamic Statistical Network Informatics, a National Science Foundation Graduate Research Fellowship (DGE 1745016), and by the center for Computational Analysis of Social and Organizational Systems (CASOS). The views and conclusions contained in this document are those of the authors and should not be interpreted as representing the official policies, either expressed or implied, of the ONR or the U.S. government.

\section*{Author's Contributions}
Gian Maria Campedelli, Iain Cruickshank and Kathleen M. Carley have developed together the theoretical setup of the study. Gian Maria Campedelli and Iain Cruickshank have created the algorithmic framework, conducted the quantitative analyses and written the paper. Kathleen M. Carley has supervised the entire project. 

\section*{Acknowledgements}
The authors wish to thanks the two anonymous reviewers for their comments and Bruce Desmarais, Cecilia Meneghini, Alberto Aziani and Pasquale De Meo for their precious suggestions on earlier versions of this manuscript.


\bibliographystyle{spbasic}
\bibliography{gow.bib}      

\end{backmatter}
\end{document}